\documentclass[preprint, 3p, sort&compress]{elsarticle}
\usepackage{dcolumn}
\usepackage{booktabs,tabularx} 
\usepackage{booktabs}
\usepackage{textcomp}
\usepackage{amssymb}
\usepackage{amsthm}
\usepackage[mathlines]{lineno}
\usepackage{graphicx}
\usepackage{units}
\usepackage{url}
\usepackage{amsmath}
\usepackage{amsfonts}
\usepackage{bm}
\usepackage{textcomp}
\usepackage{subfigure}
\usepackage{multicol}
\usepackage{multirow}
\usepackage{verbatim}
\usepackage{rotating}
\usepackage[colorlinks,linkcolor=black,citecolor=red]{hyperref}
\usepackage{float}
\usepackage{epsfig}
\usepackage{dcolumn}
\usepackage{bm}
\usepackage{color}
\usepackage{pstricks}
\usepackage{pst-node}
\usepackage{times}
\usepackage{indentfirst}
\usepackage[english]{babel}
\usepackage{epstopdf}
\addto{\captionsenglish}{%

}
\journal{Physics Letters B}


\usepackage[font=footnotesize]{caption}

\usepackage{overpic}

\newcommand{\mevcc}{\,\mbox{MeV}/c^2}

\newcommand{\gev}{\,\mbox{GeV}}

\newcommand{\gevcc}{\,\mbox{GeV}/c^2}

\newcommand{\jpsi}{J/\psi}
\newcommand{\ra}{\rightarrow}
\newcommand{\inv}{+{\rm invisible}}
\newcommand{\sigp}{\Sigma^+}
\newcommand{\sigm}{\bar{\Sigma}^-}
\newcommand{\delp}{\Delta(1232)^+}
\newcommand{\delm}{\bar{\Delta}(1232)^-}
\newcommand{\piz}{\pi^0}
\newcommand{\pip}{\pi^+}
\newcommand{\pim}{\pi^-}
\newcommand{\pb}{\bar{p}}
\newcommand{\ext}{E_{\rm extra}}

\biboptions{numbers,sort&compress}
\begin{document}
\begin{frontmatter}
\title{{\boldmath \bf Search for a massless particle beyond the Standard Model in the  $\sigp\ra p\inv$ decay}}

\author{\begin{small}
\begin{center}
M.~Ablikim$^{1}$, M.~N.~Achasov$^{4,c}$, P.~Adlarson$^{75}$, O.~Afedulidis$^{3}$, X.~C.~Ai$^{80}$, R.~Aliberti$^{35}$, A.~Amoroso$^{74A,74C}$, Q.~An$^{71,58,a}$, Y.~Bai$^{57}$, O.~Bakina$^{36}$
% ...
\\
\vspace{0.2cm}
(BESIII Collaboration)\\
\vspace{0.2cm} {\it
$^{1}$ Institute of High Energy Physics, Beijing 100049, People's Republic of China\\
% ...
$^{3}$ Bochum  Ruhr-University, D-44780 Bochum, Germany\\
$^{4}$ Budker Institute of Nuclear Physics SB RAS (BINP), Novosibirsk 630090, Russia\\
% ...
$^{35}$ Johannes Gutenberg University of Mainz, Johann-Joachim-Becher-Weg 45, D-55099 Mainz, Germany\\
$^{36}$ Joint Institute for Nuclear Research, 141980 Dubna, Moscow region, Russia\\
% ...
$^{57}$ Southeast University, Nanjing 211100, People's Republic of China\\
$^{58}$ State Key Laboratory of Particle Detection and Electronics, Beijing 100049, Hefei 230026, People's Republic of China\\
% ...
$^{71}$ University of Science and Technology of China, Hefei 230026, People's Republic of China\\
% ...
$^{74}$ University of Turin and INFN, (A)University of Turin, I-10125, Turin, Italy; (B)University of Eastern Piedmont, I-15121, Alessandria, Italy; (C)INFN, I-10125, Turin, Italy\\
$^{75}$ Uppsala University, Box 516, SE-75120 Uppsala, Sweden\\
% ...
$^{80}$ Zhengzhou University, Zhengzhou 450001, People's Republic of China\\
\vspace{0.2cm}
$^{a}$ Deceased\\
% ...
$^{c}$ Also at the Novosibirsk State University, Novosibirsk, 630090, Russia\\
% ...
}\end{center}

\vspace{2cm}
\end{small}}
\date{\today}

\begin{abstract}
A massless particle beyond the Standard Model is searched for in the two-body decay $\sigp\ra p\inv$ using $(1.0087\pm0.0044)\times10^{10}$ $\jpsi$ events collected at a center-of-mass energy of $\sqrt{s}=3.097\gev$ with the BESIII detector at the BEPCII collider. No significant signal is observed, and the upper limit on the branching fraction $\mathcal{B}(\sigp\ra p\inv)$ is determined to be $3.2\times10^{-5}$ at the 90\% confidence level. This is the first search for a flavor-changing neutral current process with missing energy in hyperon decays which plays an important role in constraining new physics models.
\begin{keyword}
BESIII \sep FCNC process\sep hyperon decay\sep BSM particle
\end{keyword}
\end{abstract}
\end{frontmatter}

\begin{multicols}{2}
\section{Introduction}

In the Standard Model (SM) of particle physics, the flavor-changing neutral current (FCNC) decay of a meson or baryon containing strange quarks  into a final state with missing energy predominantly arises from the loop-induced quark transition $s\ra d\nu\bar{\nu}$~\cite{loop,prospect}, which is strongly suppressed by the Glashow-Iliopoulos-Maiani mechanism~\cite{GIM}. The branching fractions (BFs) of such decays for hyperons are predicted by the SM to be less than $10^{-11}$~\cite{new_inv}. However, when involving contributions from new invisible particles beyond the SM, the BFs of some FCNC hyperon decays are allowed to be as high as order $10^{-4}$~\cite{dp}. The search for this category of decays is therefore  a sensitive probe for new physics (NP).

This study aims to search for a massless particle beyond the SM, such as the massless dark photon ($\gamma'$), which can lead to invisible signatures in FCNC decays. The massless dark photon is a gauge boson associated with a new unbroken $U(1)_d$ symmetry~\cite{dp_1,dp_2}. It does not directly interact with the SM fermions but could induce FCNC processes via higher-dimensional operators~\cite{dp_3}. Another example is the QCD axion ($a$), a pseudoscalar boson originally predicted as the Peccei-Quinn solution to the strong $C\!P$ problem~\cite{axion_1,axion_2}.  With a weak coupling to fermions, it could induce $s \rightarrow d$ quark transitions. The QCD axion is expected to have a mass less than an eV and a lifetime longer than the age of the universe~\cite{axion_3}, making this study relevant to it as well.

In the meson sector, there are ongoing experimental searches for $s\ra d\nu\bar{\nu}$ transition via the kaon decays $K^{+} \rightarrow \pi^{+} \nu \bar{\nu}$ and $K_L \rightarrow \pi^0 \nu \bar{\nu}$, from the NA62~\cite{NA62} and KOTO~\cite{KOTO} Collaborations, respectively. The measurements show a slight excess with respect to the SM expectations, which has led to various  NP interpretations~\cite{new_phy}.  
Studies of rare hyperon FCNC transitions offer promising opportunities to test the SM and to search for possible NP.

This Letter reports a search for a massless beyond-the-SM `invisible' particle through a missing-energy signature, in the two-body decay $\sigp\ra p\inv$, where the $\sigp$ candidate is identified by tagging a $\sigm$ decaying to $\pb\piz$ on its recoiling side~\cite{tag}.  The analysis exploits around $10^7$ $\sigp\sigm$ hyperon pairs produced from $(1.0087\pm0.0044)\times10^{10}$ $\jpsi$ decays~\cite{jpsinum} collected at a center-of-mass energy of $\sqrt{s}=3.097\gev$ with the BESIII detector at the BEPCII collider.  This is the first experimental search for an FCNC process with missing energy in hyperon decays. A semi-blind procedure is performed to avoid possible bias, where approximately 10\% of the full data set is used to validate the analysis strategy. The final result is then obtained with the full data set only after the analysis strategy has been fixed. Throughout this Letter, charge conjugation is always implied unless mentioned otherwise.

\section{BESIII detector and Monte Carlo simulation}

The BESIII detector~\cite{Ablikim:2009aa} records symmetric $e^+e^-$ collisions 
provided by the BEPCII storage ring~\cite{Yu:IPAC2016-TUYA01}
in the center-of-mass energy range from 2.0 to 4.95~GeV. 
BESIII has collected large data samples in this energy region~\cite{Ablikim:2019hff}. The cylindrical core of the BESIII detector covers 93\% of the full solid angle and consists of a helium-based
 multilayer drift chamber~(MDC), a plastic scintillator time-of-flight
system~(TOF), and a CsI(Tl) electromagnetic calorimeter~(EMC),
which are all enclosed in a superconducting solenoidal magnet
providing a 1.0~T magnetic field.
The magnetic field was 0.9~T in 2012, which affects 11\% of the total $J/\psi$ data.
The solenoid is supported by an
octagonal flux-return yoke with resistive plate counter muon
identification modules interleaved with steel. 
The charged-particle momentum resolution at $1~{\rm GeV}/c$ is
$0.5\%$, and the 
${\rm d}E/{\rm d}x$
resolution is $6\%$ for electrons
from Bhabha scattering. The EMC measures photon energies with a
resolution of $2.5\%$ ($5\%$) at $1$~GeV in the barrel (end-cap)
region. The time resolution in the TOF barrel region is 68~ps, while
that in the end-cap region is 110~ps. The end-cap TOF
system was upgraded in 2015 using multigap resistive plate chamber
technology, providing a time resolution of
60~ps, which benefits 87\% of the data used in this analysis~\cite{etof}.

Simulation samples produced with a {\sc geant4}-based~\cite{geant4} Monte Carlo (MC) package, which
includes the geometric description~\cite{geometric} of the BESIII detector and the
detector response, are used to determine detection efficiencies
and to estimate backgrounds. The simulation models the beam-energy spread and initial-state radiation in the $e^+e^-$
annihilations with the generator {\sc
kkmc}~\cite{ref:kkmc}. The inclusive MC sample includes both the production of the $J/\psi$
resonance and the continuum processes incorporated in {\sc
kkmc}. All particle decays are modelled with {\sc
evtgen}~\cite{ref:evtgen} using BFs 
either taken from the
Particle Data Group~\cite{pdg}, when available,
or otherwise estimated with {\sc lundcharm}~\cite{ref:lundcharm}. Final-state radiation
from charged final-state particles is incorporated using the {\sc
photos} package~\cite{photos}. To study the tagging efficiency of the $\sigm\ra\pb\piz$ decay, the MC sample of $\jpsi\ra\sigp(\ra {\rm anything})\sigm(\ra\pb\piz)$ is generated according to its helicity decay amplitudes as detailed in Ref.~\cite{pgamma}.
The background process of $\jpsi\ra\delp\delm$ is generated with the angular distribution of $1+\cos^2\alpha$~\cite{ref:evtgen}, where $\alpha$ is the polar angle of $\delp$ in the $\jpsi$ rest frame. The subsequent decays of $\delp\ra {\rm anything}$ and $\delm\ra\pb\piz$ are described by a uniform phase-space model. The signal process of $\jpsi\ra\sigp(\ra p\inv)\sigm(\ra\pb\piz)$ is generated according to its helicity decay amplitudes, where the decay-asymmetry parameter of  $\sigp\ra p\inv$ is assumed to be the same as that of $\sigp\ra p\gamma$ decay~\cite{pgamma}. 

\section{Event selection}
\subsection{Analysis method}

For the signal process of $\sigp\ra p\inv$, the $\sigp$ hyperon is inferred by reconstructing the $\sigm$ decay in the events of $\jpsi\ra\sigp\sigm$ at the center-of-mass energy of $\sqrt{s}=3.097\gev$. 
The $\sigm$ candidates, which constitute the single-tag (ST) sample, are reconstructed with the dominant decay $\sigm\ra \pb\piz$. Then the double-tag (DT) event is formed by reconstructing the signal decay $\sigp\ra p\inv$ in the system recoiling against the $\sigm$ hyperon. The absolute BF of the signal decay is determined by

\begin{equation}
\mathcal{B}_{\mathrm{sig}}=\frac{N_{\mathrm{DT}}^{\rm{obs}} / \epsilon_{\mathrm{DT}}}{N_{\mathrm{ST}}^{\rm{obs}} / \epsilon_{\mathrm{ST}}},
\end{equation}
where $N_{\rm{ST}}^{\rm{obs}}$ $(N_{\rm{DT}}^{\rm{obs}})$  is the observed ST (DT) yield and $\epsilon_{\mathrm{ST}}$ $\left(\epsilon_{\mathrm{DT}}\right)$ is the corresponding detection efficiency.

\subsection{ST selection}

Charged tracks detected in the MDC are required to be within a polar angle ($\theta$) range of $|\rm{cos\theta}|<0.93$, where $\theta$ is defined with respect to the $z$-axis,
which is the symmetry axis of the MDC. For each charged track, the distance of closest approach to the interaction point (IP) must be less than 10\,cm along the beam axis,  and less than 2\,cm
in the transverse plane~\cite{vxy}. Particle identification~(PID) for charged tracks combines measurements of the specific ionization energy loss (${\rm d}E/\rm{d}x$) in the MDC and the flight time in the TOF to form likelihoods $\mathcal{L}(h)~(h=p,K,\pi)$ for each hadron $h$ hypothesis. Tracks are identified as protons when the proton hypothesis has the greatest likelihood ($\mathcal{L}(p)>\mathcal{L}(K)$ and $\mathcal{L}(p)>\mathcal{L}(\pi)$).

Photon candidates are identified using showers in the EMC. The deposited energy of each shower must be more than 25~MeV in the barrel region ($|\!\cos\!\theta|< 0.80$) and more than 50~MeV in the end-cap region ($0.86 <|\!\cos\!\theta|< 0.92$). To exclude showers that originate from charged particles, the angle subtended by the shower in the EMC and the position of the closest charged track at the EMC must be greater than 10 degrees (20 degrees for $\pb$ candidates since anti-protons interact strongly with nuclei) as measured from the IP. To suppress electronic noise and showers unrelated to the event, the difference between the EMC time and the event start time is required to be within [0, 700]\,ns. The $\piz$ candidates are reconstructed with a pair of photons whose invariant mass is required to lie in the range of (115, 150)$\mevcc$. Candidates with both photons from end-cap EMC regions are rejected due to having a worse resolution. A kinematic fit constraining the invariant mass of the photon pair to the $\piz$ known mass~\cite{pdg} is performed, and the $\chi^2$ value must be less than 25 to ensure fit quality.

The $\sigm$ candidates are reconstructed with all $\pb\piz$ combinations, and the one with an invariant mass closest to the known $\sigm$ mass ($M_{\sigm}$)~\cite{pdg} is retained for further analysis. The invariant mass of $\pb\piz$ is required to satisfy $\left|M_{\bar{p} \pi^0}-M_{\bar{\Sigma}^{-}}\right|<15\mevcc$, which corresponds to approximately three times its resolution around $M_{\sigm}$. The yield of ST $\sigm$ hyperons is obtained by examining the distribution of the beam-constrained mass of $\pb\piz$, defined as
\begin{equation}
    M_{\rm{BC}} = \sqrt{E_{\rm{beam}}^2/c^4-\left|\vec{P}_{\pb\pi^0}\right|^2 / c^2},
\end{equation}
where $E_{\rm{beam}}$ is the beam energy and $\vec{P}_{\pb\pi^0}$ is the momentum of the reconstructed $\pb\piz$ combination in the $e^+e^-$ center-of-mass system. 

\begin{figure}[H]
  \begin{center}
    \begin{overpic}[width=7.5cm]{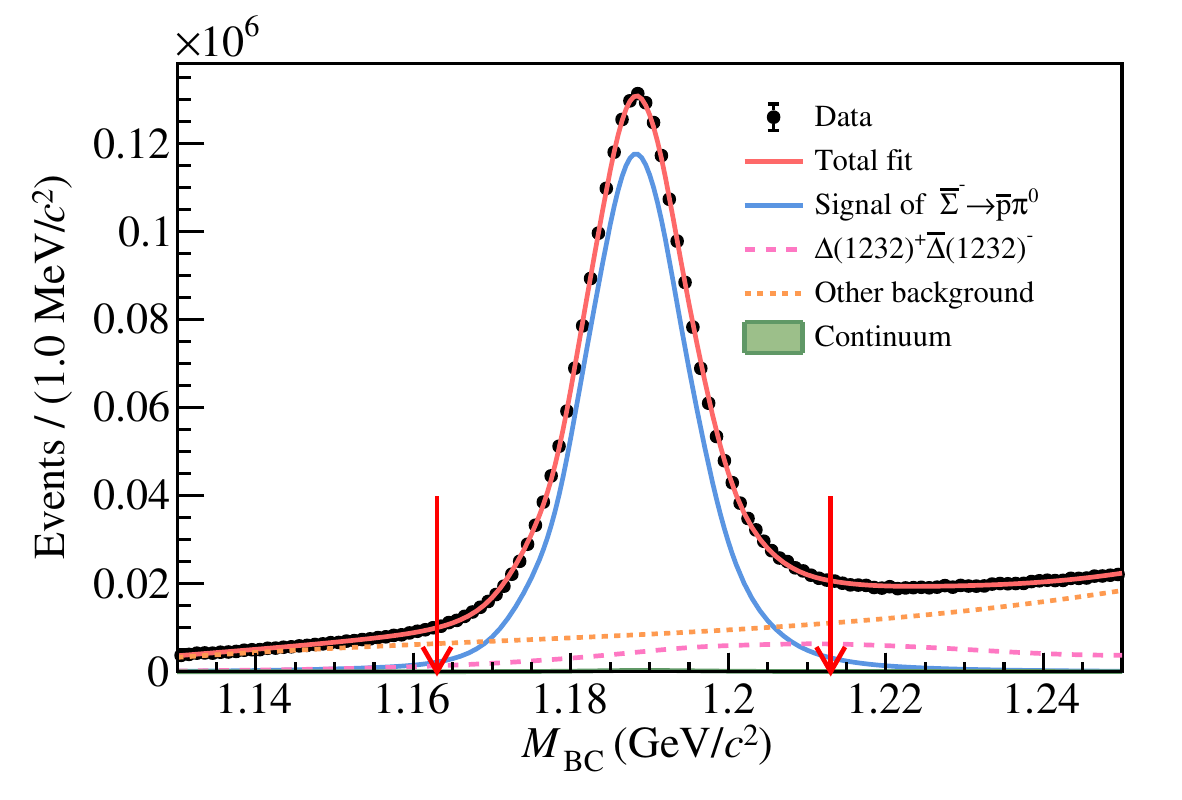}
      \put(25,50){(a)}
    \end{overpic} 
    \begin{overpic}[width=7.5cm]{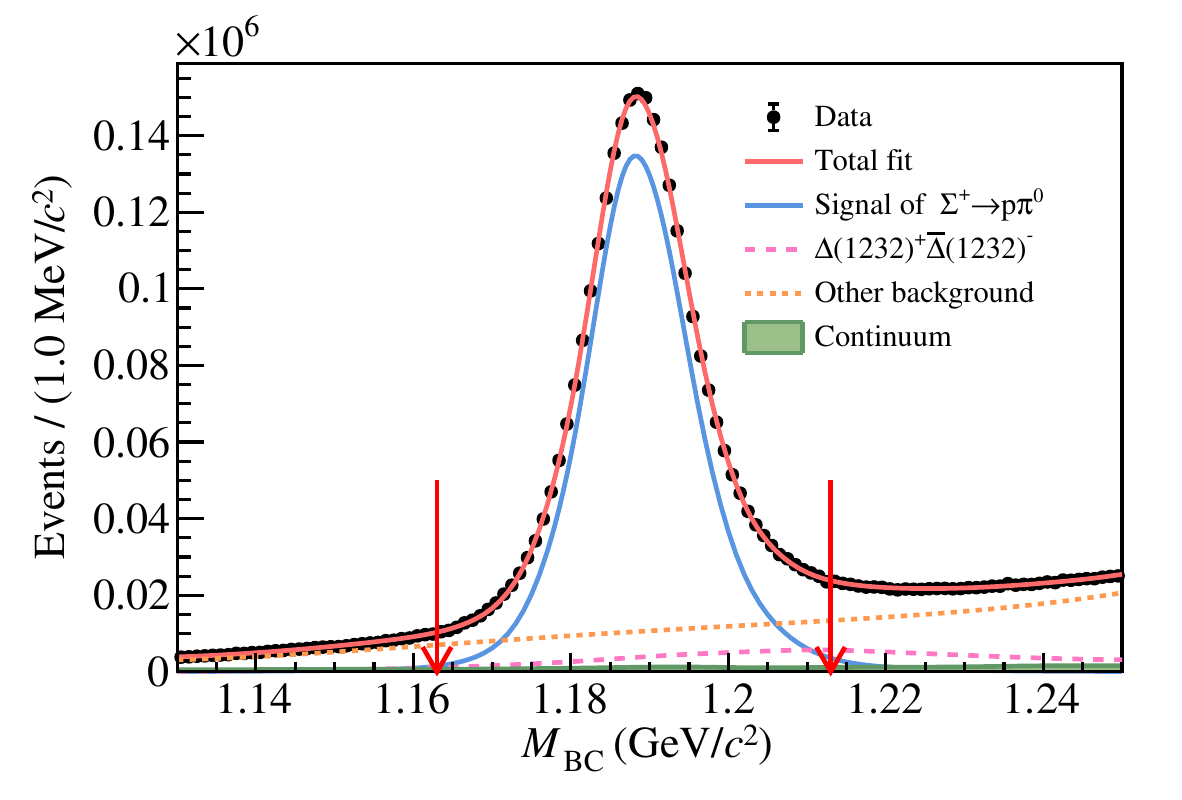}
      \put(25,50){(b)}
    \end{overpic}
    \caption{The $M_{\rm{BC}}$ distributions of ST candidates for (a) $\sigm\ra\pb\piz$ and (b) $\sigp\ra p\piz$. The red arrows indicate the ST signal windows.}
    \label{fig:fit_ST}
  \end{center}
\end{figure}

Figure~\ref{fig:fit_ST} shows the $M_{\rm{BC}}$ distributions of the ST candidates. The charge-conjugated ST candidates are reconstructed individually for each event. A binned maximum-likelihood fit is performed to the $M_{\rm{BC}}$ distribution to obtain the ST yield. In the fit, the signal and $\jpsi\ra\delp(\ra {\rm anything})\delm(\ra\pb\piz)$ background, denoted as $\delp\delm$, are described by their MC-simulated shapes convolved with a Gaussian function to account for the resolution difference between data and MC simulation. The background of the continuum processes is estimated using the data sample taken at $\sqrt{s} = 3.080\gev$ with an integrated luminosity of 168.30~pb$^{-1}$ ~\cite{jpsinum}. The yield is normalized to the $J/\psi$ data sample, taking into account the integrated luminosities and center-of-mass energies~\cite{jpsinum}. Other nonpeaking contamination, including combinatorial background, is described by a third-order Chebyshev polynomial function. The fit results are also shown in Fig.~\ref{fig:fit_ST}. The signal region is defined as $(1.163, 1.213)\gevcc$ in the $M_{\rm{BC}}$ distributions, and the ST yields of $\sigm$ and $\sigp$ are found to be $(2.077\pm0.002)\times10^6$ and $(2.356\pm0.003)\times10^6$, respectively. 

The ST detection efficiencies are evaluated using the signal MC sample, and are found to be $(37.62\pm0.04)\%$ and $(42.65\pm0.04)\%$ for $\sigm\ra\pb\piz$ and $\sigp\ra p\piz$, respectively. The BF of $\jpsi\ra\sigp\sigm$ is calculated according to the observed ST yield and the corresponding ST efficiency for the two charge-conjugated channels individually, and is found to be compatible with the previous BESIII measurement~\cite{vxy} within uncertainties.

\subsection{DT selection}

The signal process of $\sigp\ra p\inv$ is searched for using the remaining tracks recoiling against the ST $\sigm$ candidates. The following criteria are applied to select the signal candidates and suppress the backgrounds from $\jpsi\ra\sigp(\ra p\piz)\sigm(\ra\pb\piz)$ (denoted as $\sigp\ra p\piz$), $\jpsi\ra\sigp(\ra p\gamma)\sigm(\ra\pb\piz)$ (denoted as $\sigp\ra p\gamma$) and $\jpsi\ra\delp\delm$ (denoted as $\delp\delm$).
Exactly one additional charged particle has to be reconstructed for the DT candidate events and it must be identified as a proton. A two-constraint (2C) kinematic fit is performed under the hypothesis of $\jpsi\ra p\pb\piz\inv$. The fit constrains the invariant mass of two photons to the $\piz$ nominal mass and the mass of the invisible particle to zero. The $\chi^2$ value of the 2C kinematic fit ($\chi^2_{\rm 2C}$) must be less than 20 with 13 degrees of freedom. 
To suppress the $\sigp\ra p\piz$ background, another 2C kinematic fit is performed by constraining the mass of the invisible particle to the known $\piz$  mass. The obtained $\chi^2$ value ($\chi_{\rm 2C,\,\piz}^2$) is required to be larger than $\chi_{\rm 2C}^2$. 
If there are three or more photon candidates available, a five-constraint (5C) kinematic fit is performed under the hypothesis of $J / \psi \rightarrow p \bar{p} \pi^0 \gamma$ with one of all remaining photon candidates combined to the DT side, corresponding to the $\sigp$ decay process. To suppress the $\sigp\ra p\gamma$ background, each $\chi^2$ value of the 5C kinematic fit ($\chi_{\rm 5C}^2$) is required to be larger than 200 with 10 degrees of freedom. If there are four or more photon candidates available, a six-constraint (6C) kinematic fit is performed under the hypothesis of $J / \psi \rightarrow p \bar{p} \piz \gamma\gamma$ with all two-photon combinations on the DT side, where the mass of the photon pair is restricted to the known $\piz$ mass. To further suppress the $\sigp\ra p\piz$ background, each $\chi^2$ value of the 6C kinematic fit ($\chi_{\rm 6C}^2$) is required to be larger than 200 with 12 degrees of freedom. The four-momentum of the DT proton and the invisible particle is obtained from the 2C kinematic fit that constrains the mass of the invisible particle to zero. The invariant mass of the proton and the invisible particle ($M_{p+{\rm inv}}$) is required to be in the range of (1.18, 1.20)$\gevcc$.  For the $\jpsi\ra\delp\delm$ background, the final-state particles decay near the IP since the $\delp$ has a negligible lifetime compared with that of the $\sigp$ hyperon. To reduce such background, vertex fits~\cite{vertex} are performed to the $p$ and $\pb$ combination. 
The primary vertex fit constrains the tracks to originate from a common vertex, while the secondary vertex fit constrains the momentum of the $p\pb$ combination to point back to the IP. For the events passing the vertex fits, the length ($L$) from the reconstructed vertex to the IP is required to be more than twice the resolution ($\sigma_L$). The region where the polar angle of the invisible particle ($\theta_{\rm{inv}}$) in the $\jpsi$ rest frame satisfies $|\!\cos \theta_{\rm{inv}}|>0.8$ is eliminated because the $\sigp\ra p\gamma$ background predominantly lies in this region with other requirements applied according to the MC simulation. The requirements of $\chi^2_{\rm 2C}$, $M_{p+{\rm inv}}$ and $L/\sigma_L$ are optimized according to the Punzi significance~\cite{Punzi}, defined as $\varepsilon/(1.5+\sqrt{B})$, where $\varepsilon$ denotes the signal efficiency obtained from signal MC sample and $B$ is the number of background events obtained from background MC samples.

\section{DT signal extraction}

\begin{figure}[H]
	\begin{center}
		\begin{overpic}[width=7.5cm]{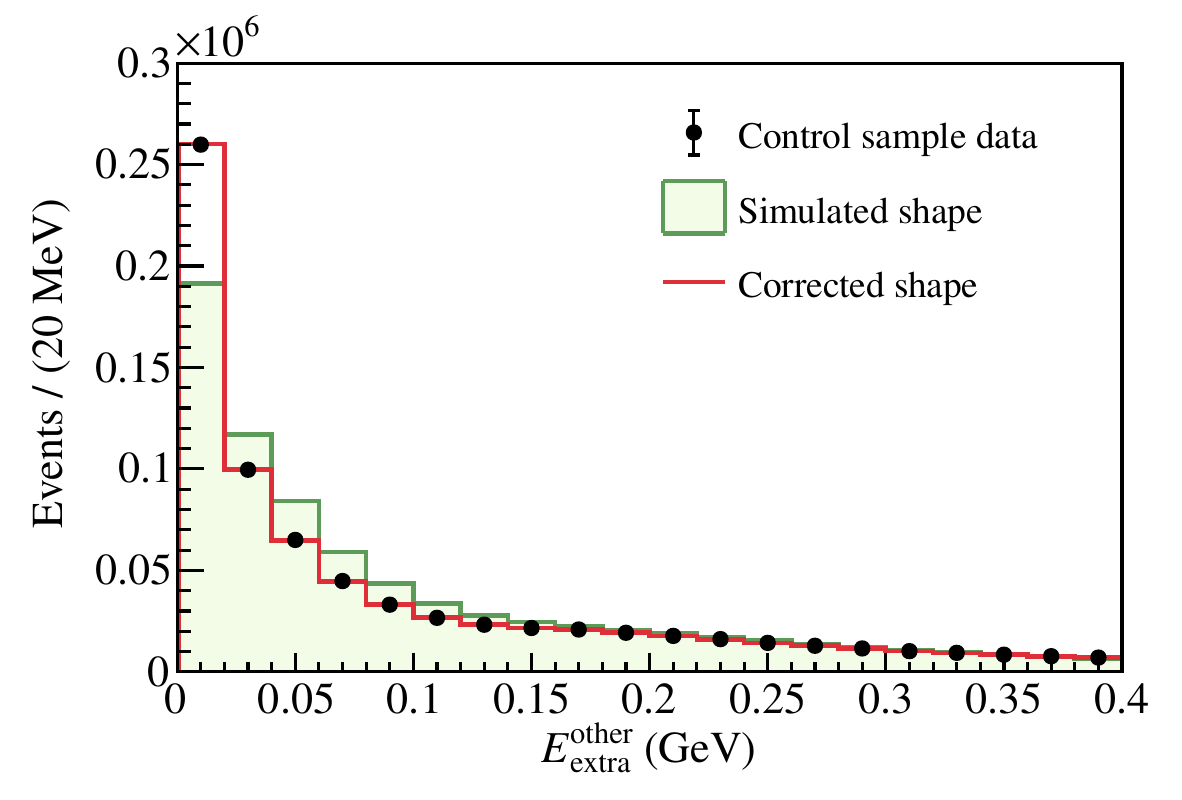}
		\end{overpic} 
  	\caption{The $\ext^{\rm{other}}$ distribution for the $\jpsi\ra\sigp(\ra p\piz)\sigm(\ra\pb\piz)$ control sample. }
		\label{fig:control}
	\end{center}
\end{figure}

After applying all the selection criteria, MC studies with a generic event type analysis tool~\cite{topoana} indicate that the dominant background events are from the $\sigp\ra p\piz$, $\delp\delm$ and $\sigp\ra p\gamma$ processes. There are additional backgrounds from other sources in the inclusive MC sample, but there is no event left from the continuum data. Since the invisible particle on the DT side does not deposit any energy in the EMC, the energy sum of all the showers in the EMC except for the ST $\piz$, $\ext$, can be utilized as a discriminator to extract the DT yield. The $\ext$ is divided into two parts
\begin{equation}
    \ext=\ext^{{\rm DT} \piz/\gamma}+\ext^{\rm{other}},
\end{equation}
where $\ext^{{\rm DT} \piz/\gamma}$ denotes the energy of the $\piz$ or $\gamma$ on the DT side, which is expected to be zero for signal events.   The value of $\ext^{{\rm DT} \pi^0/\gamma}$ in background events is obtained through the MC simulation, as the interactions of photons or electrons with the material are described in the simulation with a sufficient accuracy. 
The contribution $\ext^{\rm{other}}$ originates from other sources, including noise unrelated to the event. It is estimated that the interaction between the $\pb$ track and detector contributes to approximately 93\% of this, under the condition that the induced showers are already suppressed through the isolation angle criteria. Due to difficulties in accurately modeling anti-proton interactions with the detector material using the {\sc geant4} package, the raw simulation of $\ext^{\rm{other}}$ deviates from the data, as illustrated in Fig.~\ref{fig:control}. The shape of $\ext^{\rm{other}}$ is corrected using a data-driven approach~\cite{lambda} based on a $\jpsi\ra\sigp(\ra p\piz)\sigm(\ra\pb\piz)$ control sample. The contribution of $\ext^{\rm{other}}$  is assigned with a random value from the shape template obtained from the data control sample, according to the momentum and polar angle of the anti-proton. The corrected shape of $\ext^{\rm{other}}$ is found to have a good agreement with the control-sample data as shown in Fig.~\ref{fig:control}.

\begin{figure}[H]
  \begin{center}
    \begin{overpic}[width=7.5cm]{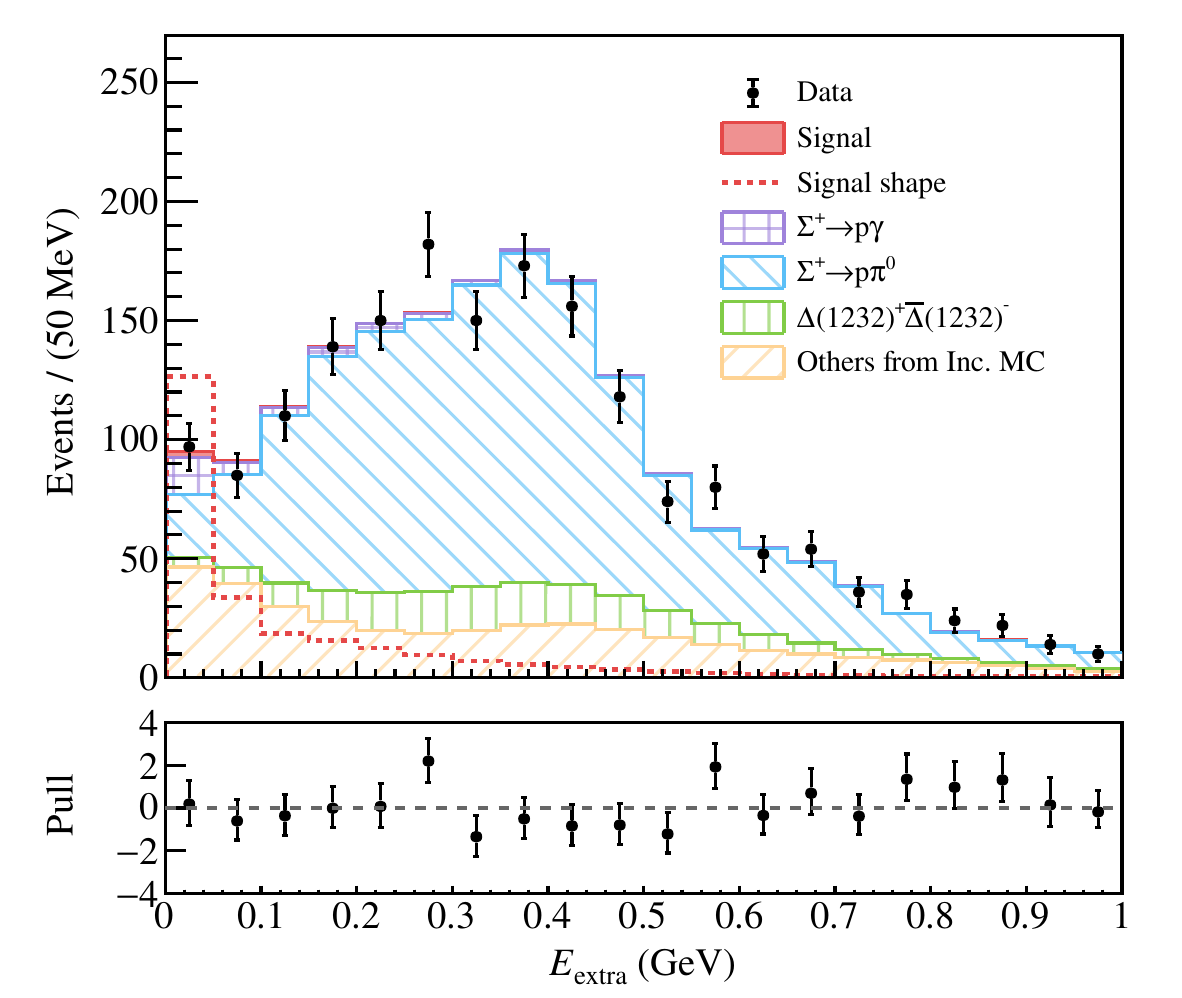}
      \put(25,70){(a)}
    \end{overpic} 
    \begin{overpic}[width=7.5cm]{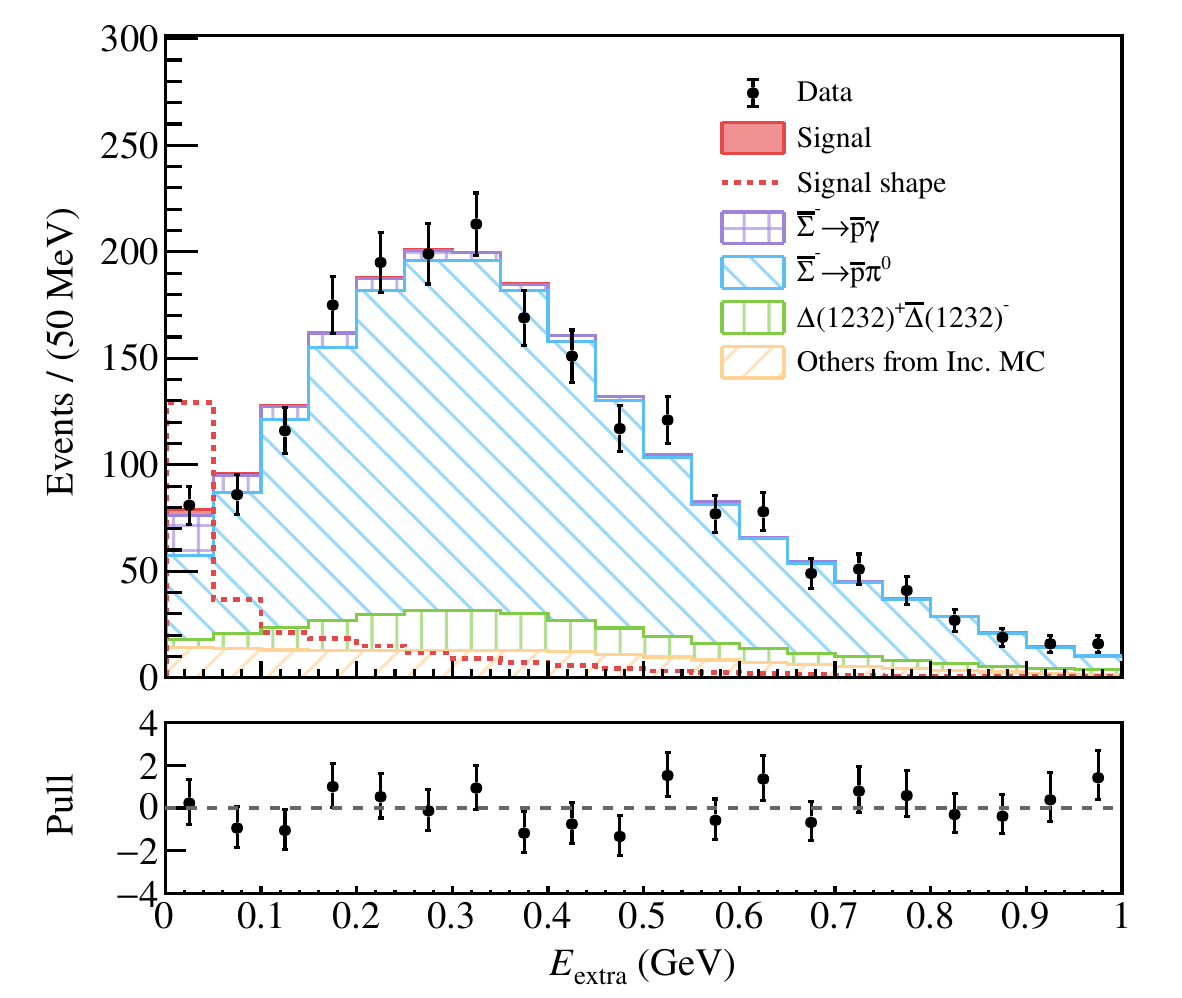}
      \put(25,70){(b)}
    \end{overpic}
    \caption{The post-fit distributions of $\ext$ for (a) $\sigp\ra p\inv$ and (b) $\sigm\ra\pb\inv$ DT signal channels. The signal shape is normalized to a BF of $3.0 \times 10^{-4}$. The bottom panel shows the fit residuals. }
    \label{fig:fit_DT}
  \end{center}
\end{figure}

The corrected distribution of $\ext$ is used as input in a binned maximum-likelihood fit to determine the DT signal yield,  performed simultaneously on the two charge-conjugated channels, assuming the same BF of the signal in both. In the fit, the signal, backgrounds of $\sigp\ra p\piz$, $\delp\delm$, $\sigp\ra p\gamma$, and other backgrounds in the inclusive MC sample are described by their MC-simulated shapes after the data-driven correction. The Gaussian process regression method \cite{GPR} is utilized to smooth the MC shapes of $\sigp\ra p\piz$ and $\delp\delm$ backgrounds. The relative ratio of the yields for the two background components is determined with a control sample of $\jpsi\ra p\pb\piz\piz$~\cite{pgamma}. The background yield of $\sigp\ra p\gamma$ is estimated using the MC sample and normalized according to the BF of $\sigp\ra p\gamma$~\cite{pgamma}. A kernel density estimation method~\cite{KDE} is used to smooth the MC shape of other backgrounds in the inclusive MC sample, with its yield normalized to the total number of $\jpsi$ events~\cite{jpsinum}.
Figure ~\ref{fig:fit_DT} shows the post-fit distributions of $\ext$. No significant signal contribution is observed. The BF of $\sigp\ra p\inv$ is determined to be $(0.6\pm1.5)\times10^{-5}$, where the uncertainty is only statistical.

\section{Systematic uncertainty}

The use of the DT technique in the analysis means that most of the systematic uncertainties related to the ST selection cancel out. The remaining systematic uncertainties are divided into two types: additive and multiplicative. The additive uncertainties are related to the specific fit methods, while the multiplicative uncertainties are associated with the knowledge of the signal efficiency.

When performing the binned maximum-likelihood fit to the $\ext$ distribution, the uncertainty arising from the choice of bin width is considered by using alternative bin widths of 40 MeV and 33 MeV. The uncertainty due to the signal shape is assigned by considering alternative signal models in which the decay asymmetry parameter of the $\sigp\ra p\inv$ decay is varied between -1 and 1. The uncertainty due to the background shape of $\sigp\ra p\piz$ and $\sigp\ra p\gamma$ decays is accounted for by varying the corresponding decay parameters within $1\sigma$~\cite{pdg,pgamma}. The uncertainty arising from the background shape of $\delp\delm$ is assessed by using an alternative phase-space model to describe the $\jpsi\ra\delp\delm$ process. The uncertainty arising from the shape of other sources of contamination in the inclusive MC sample is estimated by varying the bandwidth of the kernel function within a reasonable range. The fit is performed fourteen times in total with different methods, and the minimum significance value and the maximum upper limit are recorded.
The uncertainties of the relative ratio of the background $\sigp\ra p\piz$ and $\delp\delm$, the yield of the background $\sigp\ra p\gamma$, and the yield of other backgrounds in the inclusive MC sample are incorporated into the general likelihood assuming a Gaussian distribution. 

The multiplicative systematic uncertainties are listed in Table~\ref{tab:multi}. The uncertainty due to the ST yield (0.4\%) is evaluated by approximating all background contributions with a third-order Chebyshev polynomial function. The uncertainty due to proton tracking and particle identification (0.4\%) is studied with a $\jpsi\ra p\pb\pip\pim$ control sample. The uncertainty associated with the  $\chi_{\rm 5C}^2$ and $\chi_{\rm 6C}^2$ requirements (0.1\%) is assessed using a control sample of $\jpsi\ra\sigp(\ra p\gamma)\sigm(\ra\pb\piz)$ decays. The uncertainty of the signal model (3.6\%) is studied by varying the signal shape obtained from signal MC samples with different decay parameters. The uncertainties arising from the $\chi_{\rm 2C}^2$ (0.3\%), $\chi_{\rm 2C}^2<\chi_{\rm 2C,\,\piz}^2$ (0.1\%), $M_{p+{\rm inv}}$ (0.4\%), decay length (0.6\%) and $\cos \theta_{\rm{inv}}$ (0.3\%) requirements are assigned from studies of a control sample of $\jpsi\ra\sigp(\ra p\piz)\sigm(\ra\pb\piz)$ decays. By assuming all the sources to be independent, the total multiplicative systematic uncertainty (3.7\%) is included in the overall likelihood as a Gaussian nuisance parameter with a width equal to the uncertainty.

\begin{table}[H]
\centering
\caption{The multiplicative systematic uncertainties.}
\label{tab:multi}
\begin{tabular}{lc}
\hline Source & Uncertainty (\%)  \\
\hline ST yield & 0.4  \\
Tracking and PID & 0.4 \\
$\chi_{\rm 2C}^2$ requirement& 0.3 \\
$\chi_{\rm 2C}^2<\chi_{\rm 2C,\,\piz}^2$ & 0.1 \\
$\chi_{\rm 5C}^2$ and $\chi_{\rm 6C}^2$ requirements& 0.1 \\
$M_{p+{\rm inv}}$ requirement & 0.4 \\
Decay length requirement & 0.6 \\
$\cos \theta_{\rm{inv}}$ requirement & 0.3 \\
Signal model & 3.6 \\ \hline
Total (multiplicative) & 3.7 \\
\hline
\end{tabular}
\end{table}

\section{Result} 
\begin{figure}[H]
  \begin{center}
    \begin{overpic}[width=7.5cm]{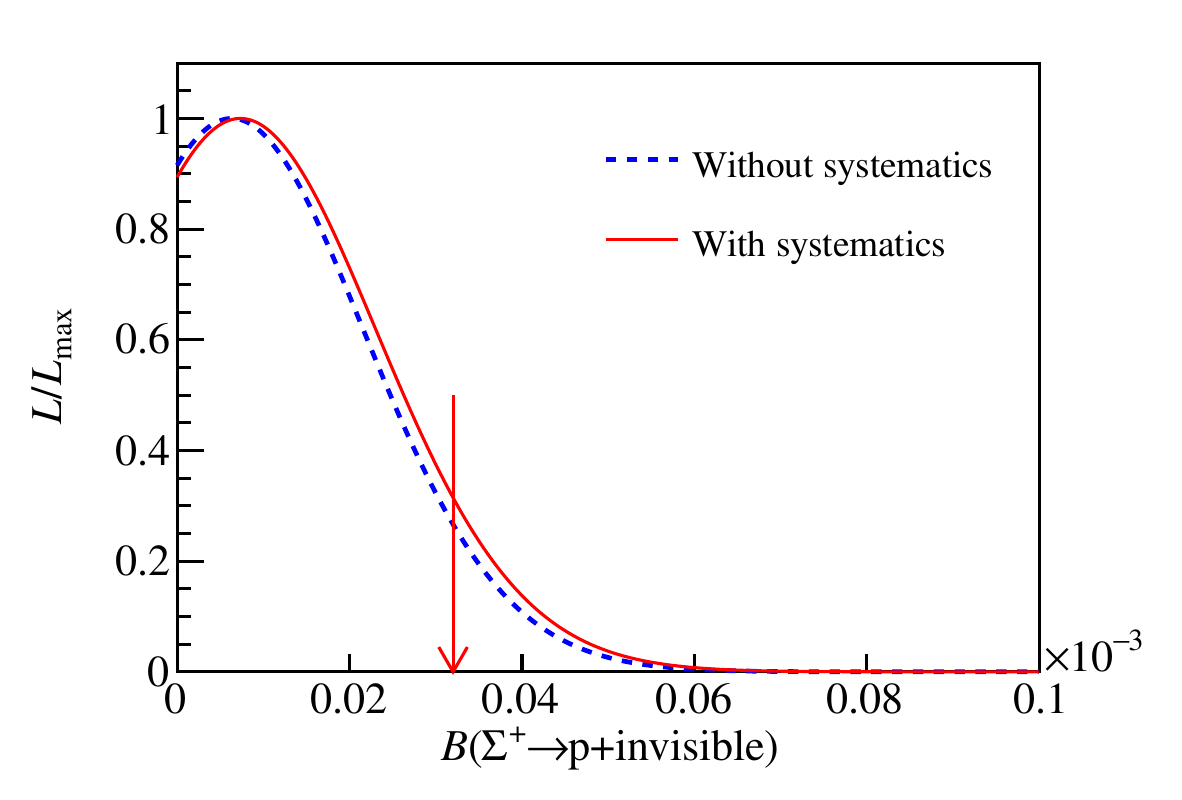}
    \end{overpic} 
    \caption{The normalized likelihood curve versus $\mathcal{B}(\sigp\ra p\inv)$. The red arrow indicates the 90\% CL upper limit.}
    \label{fig:like_curve}
  \end{center}
\end{figure}

Since no significant signal is observed in data, a Bayesian method is used to set the upper limit on the branching fraction $\mathcal{B}(\sigp\ra p\inv)$. A series of maximum-likelihood fits are performed to the $\ext$ distribution with $\mathcal{B}(\sigp\ra p\inv)$ fixed to a nonnegative scanning value. A likelihood curve $\mathcal{L}$ is constructed with these values of  $\mathcal{B}(\sigp\ra p\inv)$ as input. The normalized likelihood curves $\mathcal{L}/\mathcal{L}_{\rm{max}}$ with and without considering systematic uncertainties are shown in Fig.~\ref{fig:like_curve} and the 90\% confidence level (CL) upper limit on $\mathcal{B}(\sigp\ra p\inv)$ is found to be $3.2\times10^{-5}$, with the expected limit of $2.7_{-0.7}^{+1.1}\times10^{-5}$.

\begin{figure}[H]
  \begin{center}
    \begin{overpic}[width=7.5cm]{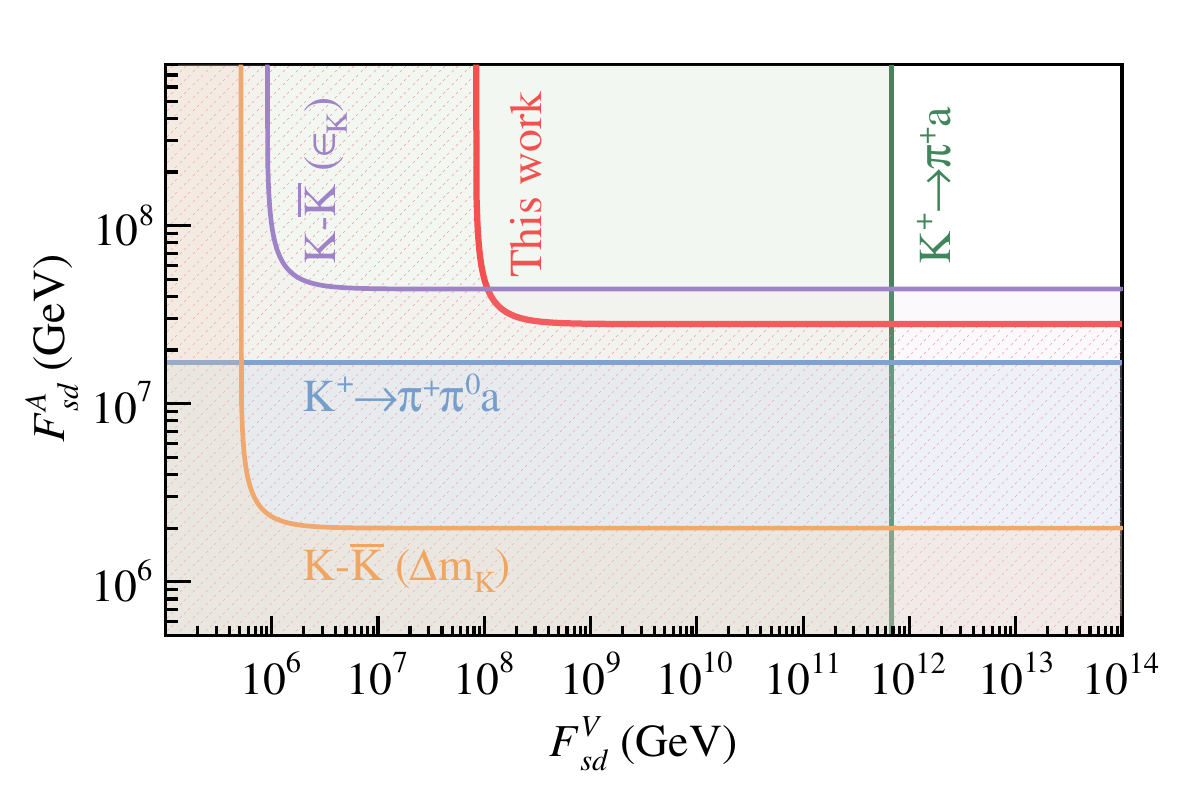}
    \end{overpic} 
    \caption{The 90\% CL exclusion limits of $s\ra d$ axion-fermion effective decay constant obtained from this analysis, where the hatched region is excluded. The $F_{sd}^{V}$ and $F_{sd}^{A}$ represent the vectorial and axial-vectorial parts of the decay constant, respectively. The constraints from $K^{+} \rightarrow \pi^{+} a$~\cite{kpi}, $K^{+} \rightarrow \pi^{+} \pi^0 a$~\cite{k2pi}, $K-\bar{K}(\Delta m_K)$~\cite{pdg}, $K-\bar{K}(\epsilon_K)$~\cite{epsilon} are also shown. When obtaining constraints from $K-\bar{K}$ mixing, the unknown low-energy constants are assumed to be zero~\cite{axion}.}
    \label{fig:couple}
  \end{center}
\end{figure}

Under the hypothesis of a massless dark photon, the maximum BF allowed for $\sigp\ra p\gamma'$ in certain scenarios is $3.8\times10^{-5}$~\cite{dp}, which lies above our upper limit. For the QCD axion, the vectorial part ($F_{sd}^{V}$) of the axion-fermion effective decay constant~\cite{axion} is highly constrained by searches for $K^+\ra\pip a$~\cite{kpi} as shown in Fig.~\ref{fig:couple}. However, for the axial-vectorial part ($F_{sd}^{A}$), a  lower bound of $F_{sd}^{A}>2.8\times10^7\gev$ is set using the upper limit obtained in this study, which is significantly better than the constraint from $K-\bar{K}$ mixing $(\Delta m_K)$~\cite{pdg} and  competitive with that from searches for $K^{+} \rightarrow \pi^{+} \pi^0 a$~\cite{k2pi} and measurements of the $C\!P$-violating parameter $\epsilon_K$ in the kaon system~\cite{epsilon}.

\section{Summary}
The first search for a massless particle beyond the SM in the two-body hyperon FCNC transition $\sigp\ra p\inv$ is presented using $(1.0087\pm0.0044)\times10^{10}$ $\jpsi$ events collected at a center-of-mass energy of $\sqrt{s}=3.097\gev$ with the BESIII detector at the BEPCII collider. No significant signal is observed and the upper limit on the branching fraction $\mathcal{B}(\sigp\ra p\inv)$ is set to be $3.2\times10^{-5}$ at the 90\% CL. This result imposes stringent limit for the NP models with a massless particle beyond the SM.

\section*{Acknowledgement}

The BESIII Collaboration thanks the staff of BEPCII and the IHEP computing center for their strong support. This work is supported in part by National Key R\&D Program of China under Contracts Nos. 2020YFA0406400, 2020YFA0406300; Joint Large-Scale Scientific Facility Funds of the NSFC and CAS under Contract No. U1832207; National Natural Science Foundation of China (NSFC) under Contracts Nos. 11635010, 11735014, 11835012, 11935015, 11935016, 11935018, 11961141012, 12005228, 12025502, 12035009, 12035013, 12061131003, 12192260, 12192261, 12192262, 12192263, 12192264, 12192265, 12221005, 12225509, 12235017; the Chinese Academy of Sciences (CAS) Large-Scale Scientific Facility Program; the CAS Center for Excellence in Particle Physics (CCEPP); CAS Key Research Program of Frontier Sciences under Contracts Nos. QYZDJ-SSW-SLH003, QYZDJ-SSW-SLH040; 100 Talents Program of CAS; The Institute of Nuclear and Particle Physics (INPAC) and Shanghai Key Laboratory for Particle Physics and Cosmology; European Union's Horizon 2020 research and innovation programme under Marie Sklodowska-Curie grant agreement under Contract No. 894790; German Research Foundation DFG under Contracts Nos. 455635585, Collaborative Research Center CRC 1044, FOR5327, GRK 2149; Istituto Nazionale di Fisica Nucleare, Italy; Ministry of Development of Turkey under Contract No. DPT2006K-120470; National Research Foundation of Korea under Contract No. NRF-2022R1A2C1092335; National Science and Technology fund of Mongolia; National Science Research and Innovation Fund (NSRF) via the Program Management Unit for Human Resources \& Institutional Development, Research and Innovation of Thailand under Contract No. B16F640076; Polish National Science Centre under Contract No. 2019/35/O/ST2/02907; The Swedish Research Council; U. S. Department of Energy under Contract No. DE-FG02-05ER41374

\end{multicols}

\end{document}